\documentclass{elsart}
\usepackage{amsmath,amssymb,graphicx}
\journal{Physica D: Nonlinear Phenomena}
\date{26 Dec 2005}
\begin{document}
\newcommand{\RN}[1]{\uppercase\expandafter{\romannumeral#1}}
\begin{frontmatter}
\title{Discrete mappings with an explicit discrete Lyapunov function related to integrable mappings}
\author[waseda]{Hironori Inoue},
\ead{hironori@ruri.waseda.jp}
\author[waseda]{Daisuke Takahashi},
\ead{daisuket@waseda.jp}
\author[ryukoku]{Junta Matsukidaira}
\ead{junta@math.ryukoku.ac.jp}
\address[waseda]{Department of Mathematical Sciences, Waseda University, 3-4-1, Okubo, Shinjuku-ku, Tokyo 169-8555, JAPAN}
\address[ryukoku]{Department of Applied Mathematics and Informatics, 
Ryukoku University, Seta, Otsu, Shiga 520-2194, JAPAN}
\begin{abstract}
  We propose discrete mappings of second order that have a discrete analogue of Lyapunov function. The mappings are extensions of the integrable Quispel-Roberts-Thompson (QRT) mapping, and a discrete Lyapunov function of the mappings is identical to an explicit conserved quantity of the QRT mapping.  Moreover we can obtain a differential and an ultradiscrete limit of the mappings preserving the existence of Lyapunov function. We also give applications of a mapping with an adjusted parameter, a probabilistic mapping and coupled mappings.
\end{abstract}
\end{frontmatter}
\section{Introduction}
  Among discrete mappings, integrable ones have a rich mathematical structure such as explicit exact solutions, conserved quantities, symmetries, and so on.  For example, most discrete soliton equations belong to the discrete KP hierarchy, have explicit $N$ soliton solutions and have an infinite number of explicit conserved quantities\cite{hirota1,hirota2,side4}.  The discrete Painlev\'e equations have a direct relation to the affine Weyl group and their mathematical structure of explicit solutions gives a hot theme\cite{noumi,kajiwara}.  Integrable systems reflect some aspects of essential mechanism of real nonlinear phenomena in conserved systems and studies on the systems are important subjects.\par
  On the other hand, general discrete systems are complex and do not exhibit such explicit mathematical structure as in the case of integrable systems. However they reflect wider nonlinear phenomena and show a rich structure of solutions, which can not be easily grasped by the integrable system theory. Thus extending the theory of integrable systems is an important theme.\par
  In this paper, we show a link between integrable and nonintegrable mappings. In section~\ref{sec:derivation}, we propose a certain type of discrete mappings that possess a discrete analogue of the Lyapunov function, which increases or decreases monotonically and converges to a certain value in the course of time evolution\cite{robinson}. We call this function a discrete Lyapunov function for short. The mappings are obtained by extending a class of second-order discrete mappings, the Quispel-Roberts-Thompson (QRT) mappings\cite{quispel1,quispel2,ramani,willox}. The QRT mappings are integrable and have an explicit conserved quantity that remains constant in the course of time evolution. The conserved quantity of QRT mappings becomes a discrete Lyapunov function in the corresponding extended mappings we propose and a trajectory of solution gets close to an attractor which is defined by the discrete Lyapunov function.\par
  In section~\ref{sec:cont and ultra}, we give a differential and an ultradiscrete limit of the mappings\cite{tokihiro,matsukidaira,grammaticos}.  The ultradiscrete equation is defined by using $\pm$ and $\max$ (or $\min$) operators.  The ultradiscretizing procedure was discovered in the integrable system theory and various discrete soliton equations can be ultradiscretized through a certain type of transformation of variables and parameters.  The differential equations and the ultradiscrete equations obtained from the mappings have also an explicit Lyapunov or discrete Lyapunov function derived from that of the mappings.\par
  In section~\ref{sec:application}, we give some applications of the mappings. Preserving monotonicity of the discrete Lyapunov functions, we can construct three types of variant of the mappings, a mapping with an adjusted parameter, a probabilistic mapping and coupled mappings.  In section~\ref{sec:remarks}, we give concluding remarks.\par
   Before going into the main subject, we make a short introduction to the QRT mapping and to the limiting procedures.  The symmetric version of QRT mapping is defined by the following second-order discrete equation\cite{quispel1,quispel2},
\begin{equation}  \label{qrt}
  x_{n+1} = \frac{f_1(x_n)-x_{n-1}f_2(x_n)}{f_2(x_n)-x_{n-1}f_3(x_n)},
\end{equation}
where $f_1(x)\sim f_3(x)$ are defined by
\begin{equation}  \label{qrt sub}
  \begin{bmatrix} f_1(x) \\ f_2(x) \\ f_3(x) \end{bmatrix}
=
  A \begin{bmatrix} x^2 \\ x \\ 1 \end{bmatrix}
\times
  B \begin{bmatrix} x^2 \\ x \\ 1 \end{bmatrix},
\end{equation}
with symmetric $3\times3$ matrices $A$ and $B$,
\begin{equation}  \label{A and B}
  A=\begin{bmatrix} a_{00} & a_{01} & a_{02} \\ a_{01} & a_{11} & a_{12} \\ a_{02} & a_{12} & a_{22} \end{bmatrix},
\qquad
  B=\begin{bmatrix} b_{00} & b_{01} & b_{02} \\ b_{01} & b_{11} & b_{12} \\ b_{02} & b_{12} & b_{22} \end{bmatrix}.
\end{equation}
Note that `$\times$' is an outer product.  If we define $h_n=h(x_{n-1},x_n)$ by
\begin{equation}
  h(x,y) = \frac{N(x,y)}{D(x,y)},
\end{equation}
where
\begin{equation}  \label{N and D}
  N(x,y)=[x^2\ x\ 1]\,A\begin{bmatrix} y^2 \\ y \\ 1 \end{bmatrix},
\qquad
  D(x,y)=[x^2\ x\ 1]\,B\begin{bmatrix} y^2 \\ y \\ 1 \end{bmatrix},
\end{equation}
we can easily show $h_{n+1}=h_n$.  Thus $h_n$ is a conserved quantity of (\ref{qrt}) and the QRT mapping is integrable.\par
  We can obtain a differential equation from the QRT mapping\cite{hirota3}.  For example, let us consider a QRT mapping
\begin{equation}  \label{qrt example}
  x_{n+1} = \frac{x_n^2+c_1x_n+c_2}{(c_3x_n+1)x_{n-1}}
\end{equation}
with a conserved quantity
\begin{equation}
  h(x_{n-1},x_n) = \frac{1}{x_{n-1}x_n}(x_{n-1}^2+x_n^2+c_1(x_{n-1}+x_n)+c_2+c_3x_{n-1}x_n(x_{n-1}+x_n)).
\end{equation}
If we use a transformation including a new parameter $\delta$ defined by
\begin{equation}
  x_n = e^{y(n\delta)}, \qquad c_j = \delta^2 \widetilde{c_j},
\end{equation}
a series expansion of (\ref{qrt example}) at $\delta\sim0$ gives
\begin{equation}
  y'' = \widetilde{c}_1e^{-y}+\widetilde{c}_2e^{-2y}-\widetilde{c}_3e^y + O(\delta).
\end{equation}
Thus we obtain a differential equation
\begin{equation}
  y'' = \widetilde{c}_1e^{-y}+\widetilde{c}_2e^{-2y}-\widetilde{c}_3e^y,
\end{equation}
as a limit equation of (\ref{qrt example}).  This equation is also integrable since it has a conserved quantity derived from $h$,
\begin{equation}
  \widetilde{h}(y,y')=\lim_{\delta\to0}\frac{1}{\delta^2}(h(e^{y(t-\delta)},e^{y(t)})-2)=(y')^2+2\widetilde{c}_1e^{-y}+\widetilde{c}_2e^{-2y}+2\widetilde{c}_3e^y.
\end{equation}
\par
  We can also obtain an ultradiscrete mapping from the QRT mapping\cite{tokihiro,matsukidaira,grammaticos}.  Consider (\ref{qrt example}) again and use a transformation of variable and parameter including a new parameter $\varepsilon$ defined by
\begin{equation}
  x_n=e^{X_n/\varepsilon},\qquad c_j = e^{C_j/\varepsilon}.
\end{equation}
Then we obtain an equation on $X_n$,
\begin{equation}
  X_{n+1} = -X_{n-1} + \varepsilon\log\frac{e^{2X_n/\varepsilon}+e^{(X_n+C_1)/\varepsilon}+e^{C_2/\varepsilon}}{e^{(X_n+C_3)/\varepsilon}+1}.
\end{equation}
Taking a limit $\varepsilon\to+0$ and using a formula
\begin{equation}  \label{ultra limit}
  \lim_{\varepsilon\to+0}\varepsilon\log(e^{A/\varepsilon}+e^{B/\varepsilon}+\cdots) = \max(A,B,\cdots),
\end{equation}
we obtain
\begin{equation}
  X_{n+1} = -X_{n-1}+\max(2X_n,X_n+C_1,C_2)-\max(X_n+C_3,0).
\end{equation}
This equation is also integrable since it has a conserved quantity derived from $h$,
\begin{equation}
\begin{aligned}
  H(X_{n-1},X_n)
 & = \lim_{\varepsilon\to+0}\varepsilon\log h(e^{X_{n-1}/\varepsilon},e^{X_n/\varepsilon}) \\
 & = \max(X_{n-1}+X_n+\max(X_{n-1},X_n)+C_3,\,2X_{n-1},\,2X_n,\\
 & \qquad\qquad \max(X_{n-1},X_n)+C_1,\,C_2)-X_{n-1}-X_n.
\end{aligned}
\end{equation}
\section{Derivation of mapping}  \label{sec:derivation}
  In this paper, we discuss discrete mappings which have the following form,
\begin{equation}  \label{general attractor}
  x_{n+1} = \frac{f_1(x_n)-x_{n-1}f_2(x_n)-\alpha_n x_{n-1}(h_n-h_\infty)}{f_2(x_n)-x_{n-1}f_3(x_n)-\alpha_n(h_n-h_\infty)},
\end{equation}
where $f_1(x)\sim f_3(x)$ and $h_n$ are the same as those defined in (\ref{qrt sub})$\sim$(\ref{N and D}) and $h_\infty$ is an arbitrary constant.  Let us consider a solution $x_n$ for $n\ge0$ from initial data $x_0$ and $x_1$ without loss of generality.  Condition for $\alpha_n$ is shown below.  Noting
\begin{equation}
\begin{aligned}
  & h_{n+1}-h_n \\
  & = -\frac{(x_{n+1}-x_{n-1})\big(x_{n+1}(f_2(x_n)-x_{n-1}f_3(x_n))-(f_1(x_n)-x_{n-1}f_2(x_n)\big)}{D(x_{n-1},x_n)D(x_n,x_{n+1})},
\end{aligned}
\end{equation}
and using (\ref{general attractor}), we obtain
\begin{equation}  \label{recurrence of h}
  h_{n+1}-h_\infty=\kappa_n(h_n-h_\infty),
\end{equation}
where
\begin{equation}  \label{beta}
  \kappa_n = 1-\alpha_n\frac{(x_{n+1}-x_{n-1})^2}{D(x_{n-1},x_n)D(x_n,x_{n+1})}.
\end{equation}
Therefore, if $\epsilon$ is a small positive constant and $0\le\kappa_n<1-\epsilon$ holds for any $n$, then $h_n$ decreases or increases monotonically and converges to $h_\infty$.  This condition is rewritten by
\begin{equation}  \label{condition}
  \epsilon< \alpha_n\frac{(x_{n+1}-x_{n-1})^2}{D(x_{n-1},x_n)D(x_n,x_{n+1})} \le 1.
\end{equation}
If this is satisfied, any trajectory of solution gets close to an attractor $h_n=h_\infty$ in a phase plane and $h_n$ becomes a discrete analogue to Lyapunov function of a differential equation.\par
  If we solve an inequality (\ref{condition}) on $\alpha_n$ with (\ref{general attractor}), a general condition of $\alpha_n$ is expressed by $x_{n-1}$ and $x_n$.  However, since (\ref{general attractor}) itself includes $\alpha_n$ and a sign of $D(x_{n-1},x_n)D(x_n,x_{n+1})$ depends on $x_{n-1}$, $x_n$ and $x_{n+1}$, the condition becomes quite complex.  Therefore, we choose special matrices $A$ or $B$ in (\ref{A and B}) and propose a sufficient condition of $\alpha_n$ to make the mapping (\ref{general attractor}) rational.\par
\begin{figure}
\begin{center}
\begin{tabular}{cc}
\includegraphics[scale=0.68]{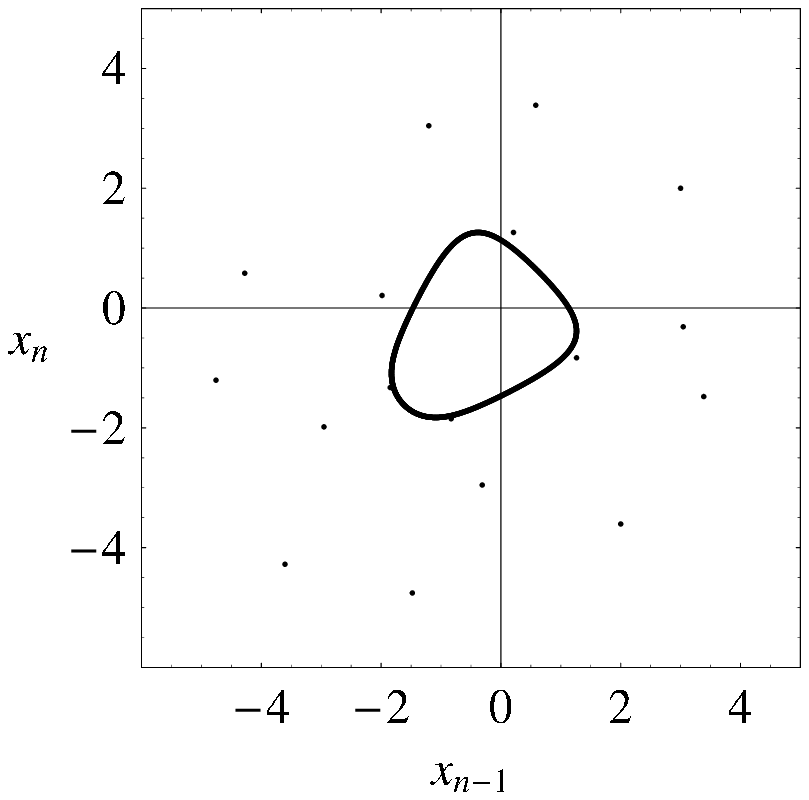} & \includegraphics[scale=0.7]{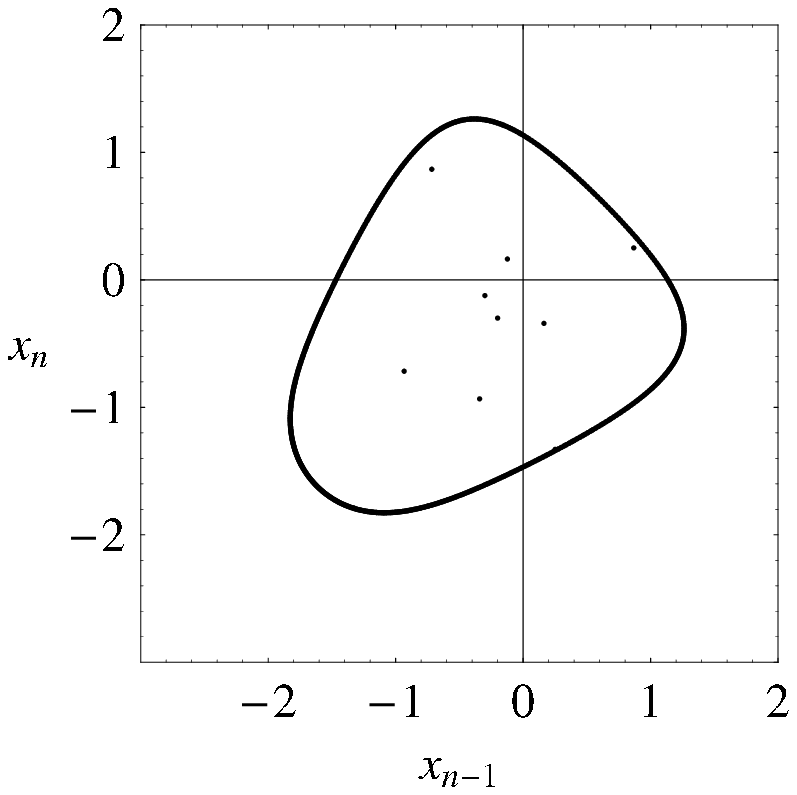} \\
(a) & (b)
\end{tabular}
\end{center}
\caption{Solutions to (\ref{class1}) with $\alpha_n$ defined by (\ref{class1 alpha}) for $(a_{00},a_{01},a_{02},a_{11},a_{12},a_{22})=(1,2,3,1,1,0)$, $h_\infty=5$ and $p=1$. Points $(x_{n-1},x_n)$ ($1\le n\le 1000$) in the phase plane are plotted.  (a) $(x_0,x_1)=(3,2)$, (b) $(x_0,x_1)=(-0.2,-0.3)$.}  \label{fig:class1 p=1}
\end{figure}

  If we choose a matrix $B$ as
\begin{equation}  \label{class1 B}
  B=\begin{bmatrix} \ 0\ &\ 0\ &\ 0\ \\\ 0\ &\ 0\ &\ 0\ \\\ 0\ &\ 0\ &\ 1\ \end{bmatrix},
\end{equation}
then $f_j$, $D$ and $h$ become as follows:
\begin{equation}
\begin{aligned}
  & f_1(x)=a_{01}x^2+a_{11}x+a_{12},\\
  & f_2(x)=-(a_{00}x^2+a_{01}x+a_{02}),\\
  & f_3(x)=0, \\
  & D(x,y)=1, \\
  & h(x,y)=a_{00}x^2y^2+a_{01}xy(x+y) \\
  & \qquad\qquad\qquad +a_{02}(x^2+y^2)+a_{11}xy+a_{12}(x+y)+a_{22}.
\end{aligned}
\end{equation}
In this case, (\ref{qrt}) is of class \RN{1} after Refs.~\cite{ramani,willox} and (\ref{general attractor}) becomes
\begin{equation}  \label{class1}
  x_{n+1} - x_{n-1} = \frac{f_1(x_n)-2x_{n-1}f_2(x_n)}{f_2(x_n)-\alpha_n(h_n-h_\infty)}.
\end{equation}
The condition (\ref{condition}) is reduced to
\begin{equation}
  \epsilon < \alpha_n \Big(\frac{f_1(x_n)-2x_{n-1}f_2(x_n)}{f_2(x_n)-\alpha_n(h_n-h_\infty)}\Big)^2 \le 1.
\end{equation}
The latter inequality gives
\begin{equation}  \label{condition2}
  \alpha_n\big((f_1(x_n)-2x_{n-1}f_2(x_n))^2+2f_2(x_n)(h_n-h_\infty)\big)\le f_2(x_n)^2+\alpha_n^2(h_n-h_\infty)^2.
\end{equation}
Then we can derive a sufficient condition for $\alpha_n$,
\begin{equation}  \label{class1 condition}
  \epsilon<\alpha_n\le\frac{f_2(x_n)^2}{(f_1(x_n)-2x_{n-1}f_2(x_n))^2+f_2(x_n)^2+(h_n-h_\infty)^2},
\end{equation}
using $2f_2(x_n)(h_n-h_\infty)\le f_2(x_n)^2+(h_n-h_\infty)^2$ and $0\le\alpha_n^2(h_n-h_\infty)^2$.  Figure~\ref{fig:class1 p=1} shows solutions to (\ref{class1}) plotted in the phase plane using $\alpha_n$ defined by
\begin{equation}  \label{class1 alpha}
  \alpha_n = p\frac{f_2(x_n)^2}{(f_1(x_n)-2x_{n-1}f_2(x_n))^2+f_2(x_n)^2+(h_n-h_\infty)^2},
\end{equation}
with an additional parameter $p$ satisfying $0<p\le1$.  Note that $f_2(x)^2=(x^2+2x+3)^2\ge4$ and the condition $\epsilon<\alpha_n$ in (\ref{class1 condition}) is satisfied in this case.
The only fixed point $x_n\sim-0.1649$ exists and does not depend on $p$.  Figure~\ref{fig:class1 p=1} (a) shows a solution to (\ref{class1}) with $p=1$ and $(x_0,x_1)=(3,2)$.  The trajectory rapidly gets close to an attractor $h_n=h_\infty$ in the outer region.  Parameters of Fig.~\ref{fig:class1 p=1} (b) are all the same as of Fig.~\ref{fig:class1 p=1} (a) other than $(x_0,x_1)=(-0.2,-0.3)$.  This trajectory moves in the inner region of the attractor and also approaches it rapidly.\par
\begin{figure}
\begin{center}
\begin{tabular}{cc}
\includegraphics[scale=0.68]{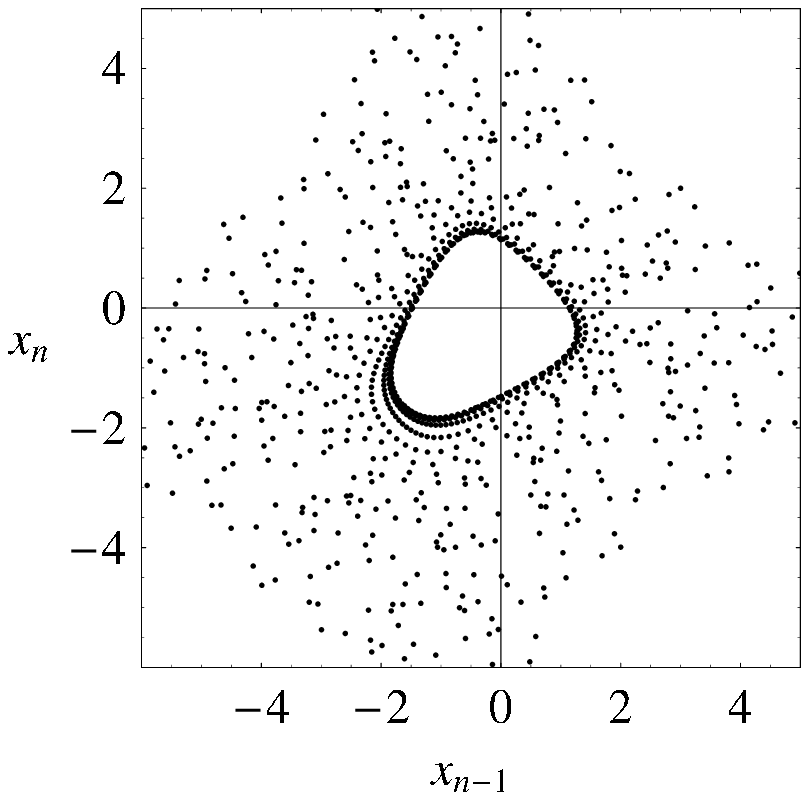} & \includegraphics[scale=0.7]{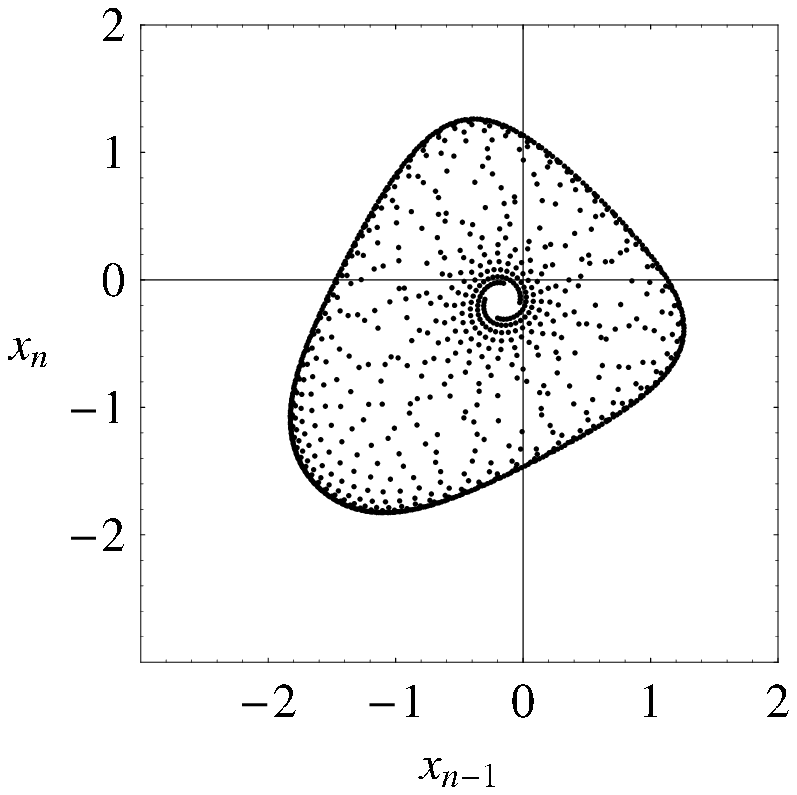} \\
(a) & (b)
\end{tabular}
\end{center}
\caption{Solutions to (\ref{class1}) with $\alpha_n$ defined by (\ref{class1 alpha}) for $p=0.02$.  Parameters other than $p$ are the same as in Fig.~\ref{fig:class1 p=1}.}  \label{fig:class1 p=0.02}
\end{figure}

\begin{figure}
\begin{center}
\begin{tabular}{cc}
\includegraphics[scale=0.7]{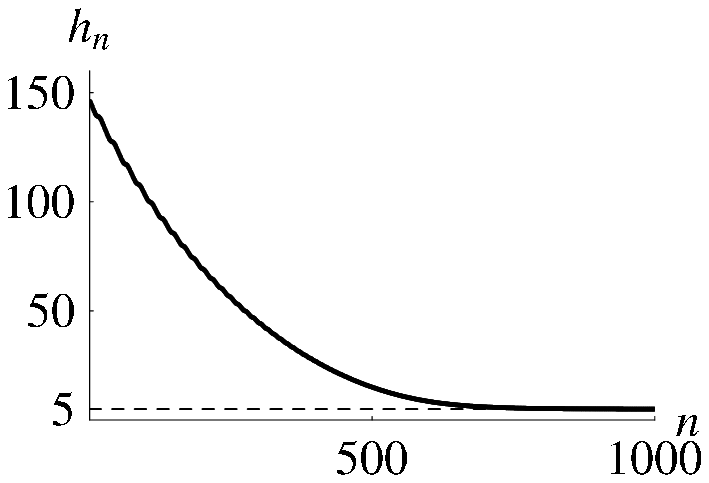} & \includegraphics[scale=0.7]{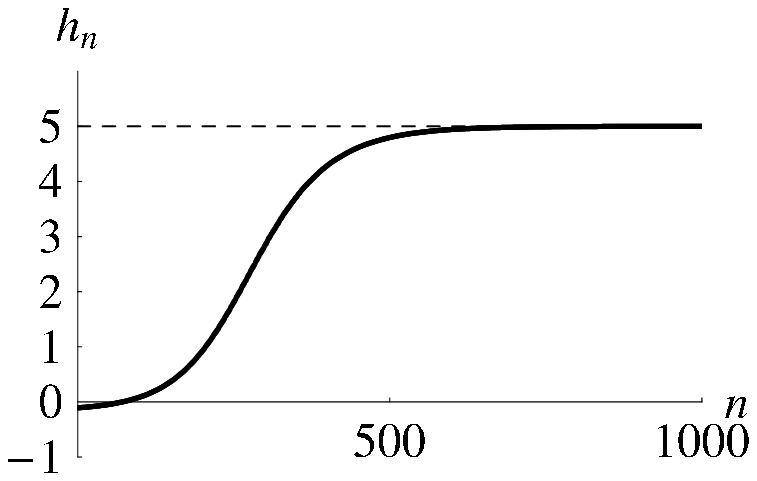} \\
(a) & (b)
\end{tabular}
\end{center}
\caption{Evolution of $h_n$ for Figs.~\ref{fig:class1 p=0.02} (a) and (b).}  \label{fig:class1 h}
\end{figure}

  If $\kappa_n=0$ in (\ref{recurrence of h}), $h_{n+1}=h_\infty$ holds and points $(x_{n-1},x_n)$ for $2\le n$ are all on the attractor $h_n=h_\infty$.  If $\kappa_n=1$, $h_{n+1}=h_n$ holds and all points are on an integrable trajectory defined by initial data and do not get close to the attractor.  Therefore, a closing rate of solution to the attractor depends on $\kappa_n$, that is, $\alpha_n$.  From the relation (\ref{beta}) between $\alpha_n$ and $\kappa_n$, we see that the closing rate is smaller if $\alpha_n$ is smaller under (\ref{class1 condition}).  Figure~\ref{fig:class1 p=0.02} shows solutions in the case of $p=0.02$ with the same parameters as in Fig.~\ref{fig:class1 p=1} other than $p$.  Since $\alpha_n$ is smaller than that of Fig.~\ref{fig:class1 p=1}, the trajectory gets close to the attractor more slowly.  Figures~\ref{fig:class1 h} (a) and (b) show an evolution of $h_n$ for Figs.~\ref{fig:class1 p=0.02} (a) and (b) respectively.  In both figures $h_n$ converges to $h_\infty$ ($=5$) monotonically.
\section{Differential and ultradiscrete equations obtained from discrete mappings}  \label{sec:cont and ultra}
\begin{figure}
\begin{center}
  \includegraphics[scale=0.7]{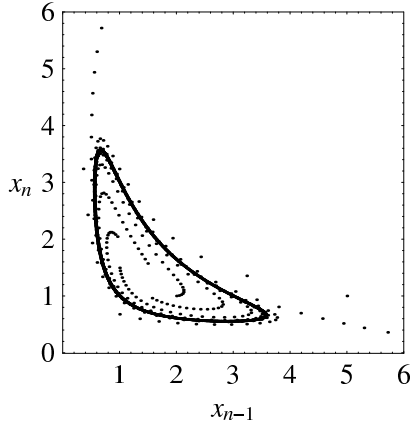}
\end{center}
\caption{Solutions to (\ref{class2}) from $(x_0,x_1)=(5,1)$ and $(2,1)$ with $(a_{00},a_{01},a_{02},a_{11},a_{12},a_{22})=(1,0,0,0,1,2)$, $h_\infty=6$, $\beta_n=100$.  Points $(x_{n-1},x_n)$ ($1\le n\le1000$) in the phase plane are plotted.} \label{fig:class2}
\end{figure}

  In this section, we consider a special case of the mapping (\ref{general attractor}) by choosing $B$ in (\ref{A and B}) as
\begin{equation}  \label{class2 B}
  B=\begin{bmatrix} \ 0\ &\ 0\ &\ 0\ \\\ 0\ &\ 1\ &\ 0\ \\\ 0\ &\ 0\ &\ 0\ \end{bmatrix}.
\end{equation}
Then $f_j$, $D$ and $h$ are given by
\begin{equation}
\begin{aligned}
  & f_1(x)=-x(a_{02}x^2+a_{12}x+a_{22}),\\
  & f_2(x)=0,\\
  & f_3(x)=x(a_{00}x^2+a_{01}x+a_{02}), \\
  & D(x,y)=xy, \\
  & h(x,y)=\frac{1}{xy}(a_{00}x^2y^2+a_{01}xy(x+y) \\
  & \qquad\qquad\qquad +a_{02}(x^2+y^2)+a_{11}xy+a_{12}(x+y)+a_{22}).
\end{aligned}
\end{equation}
In this case, (\ref{qrt}) is of class \RN{2} after Refs.~\cite{ramani,willox} and (\ref{general attractor}) becomes
\begin{equation}  \label{class2}
  x_{n+1} = \frac{g_1(x_n)}{x_{n-1}g_3(x_n)}\cdot\frac{h_\infty g_1(x_n)+h_nx_{n-1}^2g_3(x_n)+\beta_n}{h_ng_1(x_n)+h_\infty x_{n-1}^2g_3(x_n)+\beta_n},
\end{equation}
where
\begin{equation}  \label{class2 funcs}
\begin{aligned}
  g_1(x)&=-f_1(x)/x=a_{02}x^2+a_{12}x+a_{22}, \\
  g_3(x)&=f_3(x)/x=a_{00}x^2+a_{01}x+a_{02}, \\
  \alpha_n &= \frac{x_{n-1}x_ng_1(x_n)g_3(x_n)}{h_\infty(g_1(x_n)+x_{n-1}^2g_3(x_n))+\beta_n}.
\end{aligned}
\end{equation}
Note that the mapping
\begin{equation}
  x_{n+1} = \frac{g_1(x_n)}{x_{n-1}g_3(x_n)},
\end{equation}
is the original QRT mapping of class \RN{2}.  Assume
\begin{equation}  \label{class2:condition}
\begin{aligned}
 \text{(\romannumeral1)}\ & \text{$a_{ij}\ge0$ for any $i$ and $j$},\\
 \text{(\romannumeral2)}\ & \text{if $x>0$ and $y>0$, then $g_1(x)>0$, $g_3(x)>0$ and $h(x,y)>0$},\\
 \text{(\romannumeral3)}\ & \text{$h_\infty>0$ and $\beta_n\ge0$}.
\end{aligned}
\end{equation}
Then, if $x_{n-1}>0$ and $x_n>0$, we obtain $x_{n+1}>0$.  Therefore, $x_n$ from positive initial data is always positive.  If we assume that $x_n$ is always positive and that (\ref{class2:condition}) is satisfied, we can prove (\ref{condition}) is also satisfied.  The proof is complicated and is omitted here.  Under this assumption, any trajectory of solution other than a fixed point gets close to the attractor $h_n=h_\infty$ in the phase plane for any positive $\beta_n$.   Figure~\ref{fig:class2} shows examples of solution to (\ref{class2}).\par
  Moreover, there is a remarkable feature about the mapping (\ref{class2}).  Differential and ultradiscrete (piecewise-linear) equations are obtained by two kinds of limiting procedure.
\subsection{Differential equation obtained from (\ref{class2})}
\begin{figure}
\begin{center}
\begin{tabular}{cc}
  \includegraphics[scale=0.8]{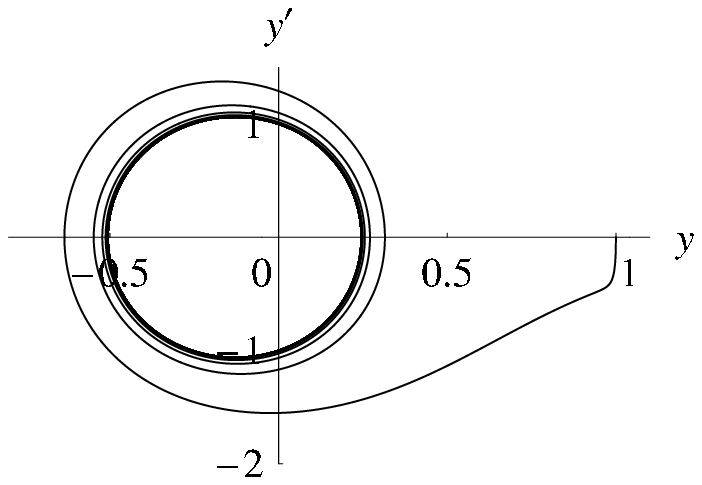} &
  \includegraphics[scale=0.8]{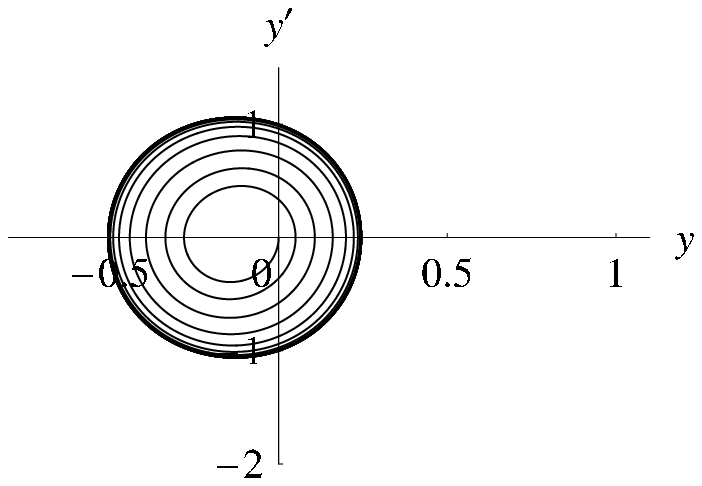} \\
  (a) & (b)
\end{tabular}
\end{center}
\caption{Solutions to (\ref{differential}) for $(c_{00}, c_{01}, c_{02}, c_{11}, c_{12}, c_{22})=(2,1,1,0,1,1)$, $\widetilde{h}_\infty=8$, $\widetilde{\beta}=0.2$ and (a) $y(0)=1$, $y'(0)=0$, (b) $y(0)=y'(0)=0$.  Both solutions are calculated numerically by the fourth-order Runge-Kutta method.}
\label{fig:differential}
\end{figure}

  With regard to a differential equation, we use the following transformation including a small parameter $\delta$.
\begin{equation}  \label{cont trans}
\begin{aligned}
  x_n &= e^{y(n\delta + \xi_0)},\\
  a_{02} &= c_{02}, & a_{11} &= -2c_{02} + \delta^2 c_{11},\\
  a_{00} &=\delta^2\,c_{00}, & a_{01} &=\delta^2\,c_{01},\qquad a_{12}=\delta^2\,c_{12},\qquad a_{22}=\delta^2\,c_{22},\\
  h_\infty &= \delta^2\,\widetilde{h}_\infty, &\beta_n &= \delta/\widetilde{\beta}.
\end{aligned}
\end{equation}
Substituting (\ref{cont trans}) to (\ref{class2}) and taking a limit $\delta\to0$, we obtain
\begin{equation}  \label{differential}
  y'' = \frac{1}{c_{02}}\Bigl(-c_{00}e^{2y}-c_{01}e^{y}+c_{12}e^{-y}+c_{22}e^{-2y}\Bigr) - 2c_{02}\,\widetilde{\beta}\,e^{2y}y'(\widetilde{h}(y,y')-\widetilde{h}_\infty),
\end{equation}
where $\widetilde{h}(y,y')$ is defined by
\begin{equation}
  \widetilde{h}(y,y')=c_{02}(y')^2+c_{00}e^{2y}+2c_{01}e^y+c_{11}+2c_{12}e^{-y}+c_{22}e^{-2y}.
\end{equation}
Note that $\widetilde{h}(y,y')$ can also be obtained from the discrete Lyapunov function $h_n=h(x_{n-1},x_n)$ as
\begin{equation}
  \widetilde{h}(y,y')=\lim_{\delta\to0}\frac{1}{\delta^2}h(e^{y((n-1)\delta+\xi_0)},e^{y(n\delta+\xi_0)}).
\end{equation}
Using (\ref{differential}), we can show
\begin{equation}
  \frac{d}{dt}(\widetilde{h}-\widetilde{h}_\infty)
= -4c_{02}^2\,\widetilde{\beta}\,e^{2y}(y')^2(\widetilde{h}-\widetilde{h}_\infty).
\end{equation}
This means $\widetilde{h}$ converges to $\widetilde{h}_{\infty}$ and $\widetilde{h}$ is a Lyapunov function to (\ref{differential}).  Figure~\ref{fig:differential} shows examples of solutions in a phase plane $(y,y')$.  The trajectories become a spiral shape approaching to a limit cycle defined by $\widetilde{h}=\widetilde{h}_\infty$.
\subsection{Ultradiscrete equation obtained from (\ref{class2})}
\begin{figure}
\begin{center}
\begin{tabular}{cc}
  \includegraphics[scale=0.7]{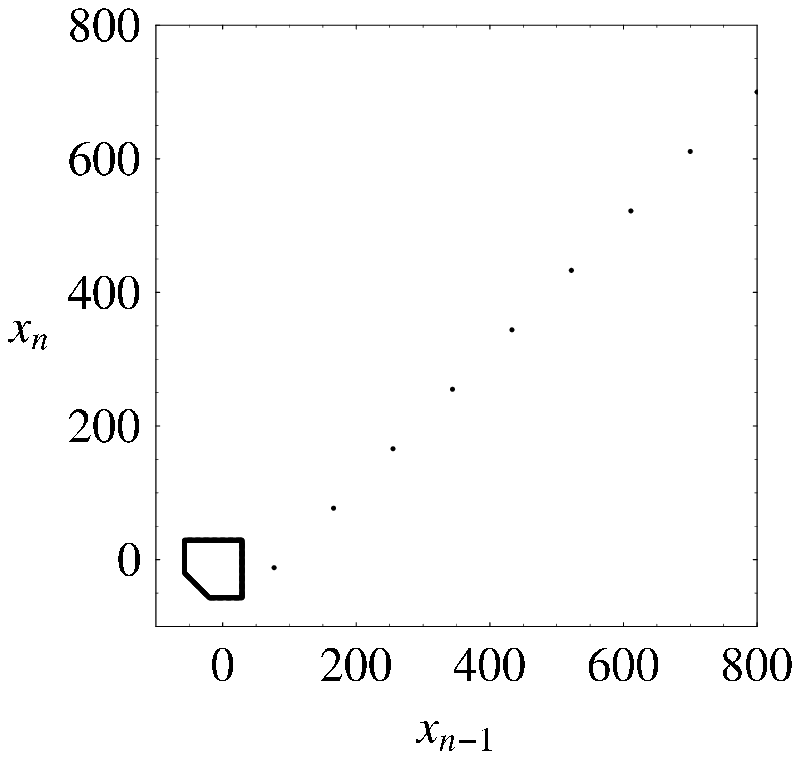} &
  \includegraphics[scale=0.7]{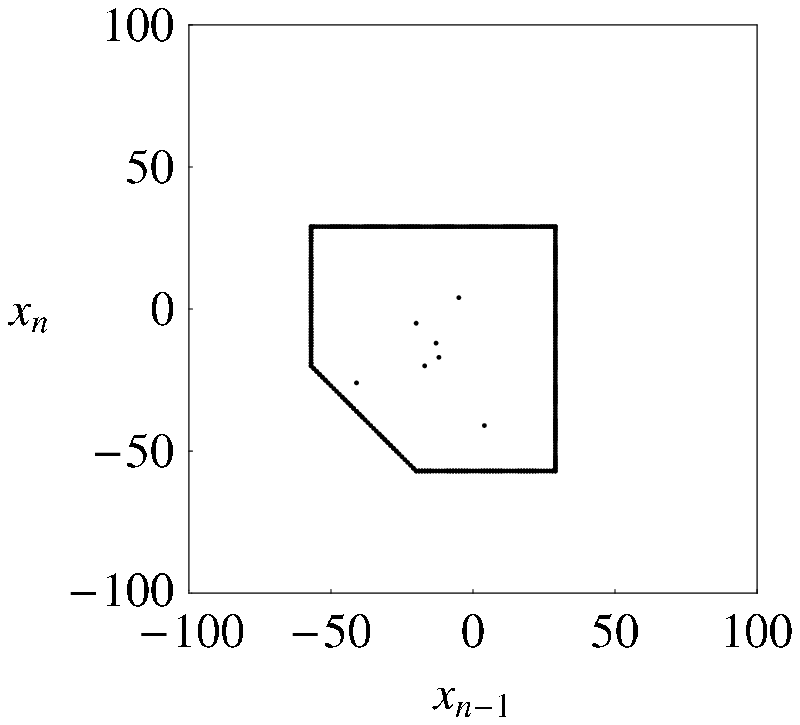} \\
  (a) & (b)
\end{tabular}
\end{center}
\caption{Solutions to (\ref{ultra}) with $(A_{00},A_{01},A_{02},A_{11},A_{12},A_{22})=(13,71,11,17,43,23)$, $H_\infty=100$. (a) $(x_0,x_1)=(800,700)$, (b) $(-13,-12)$.} \label{fig:class2 ultra1}
\end{figure}

\begin{figure}
\begin{center}
\begin{tabular}{ccc}
  \includegraphics[scale=0.5]{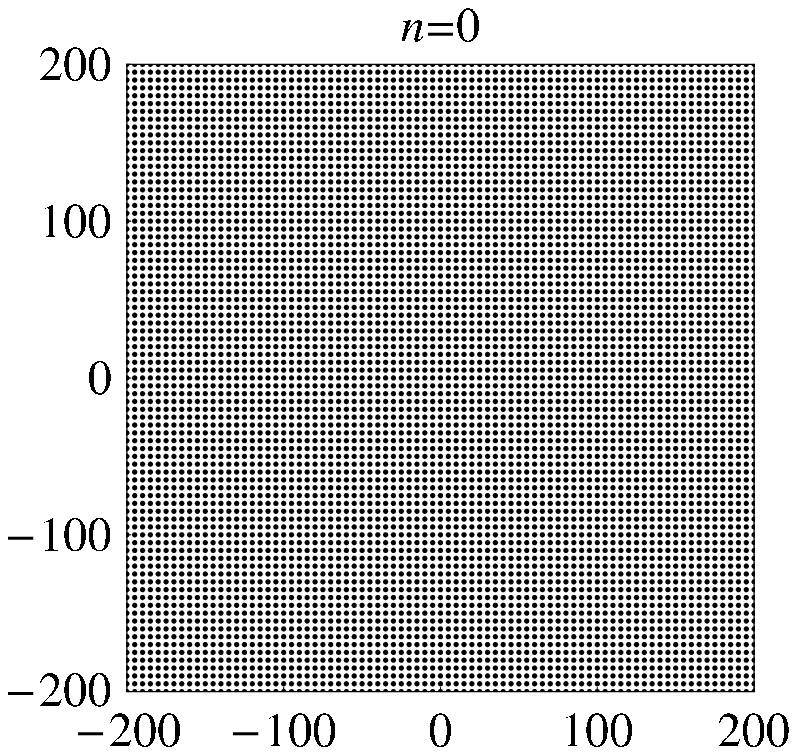} &
  \includegraphics[scale=0.5]{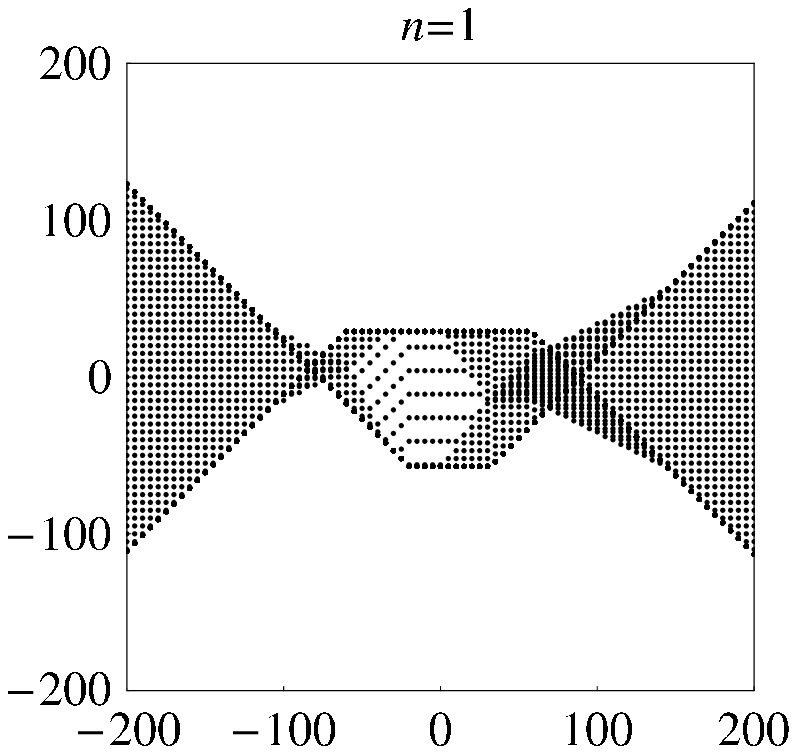} &
  \includegraphics[scale=0.5]{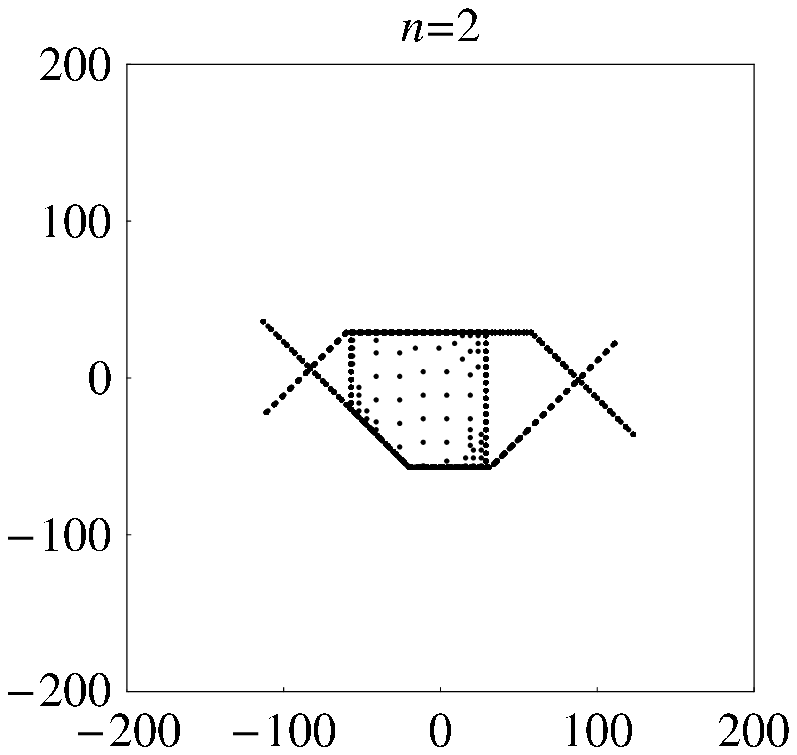} \\
  \includegraphics[scale=0.5]{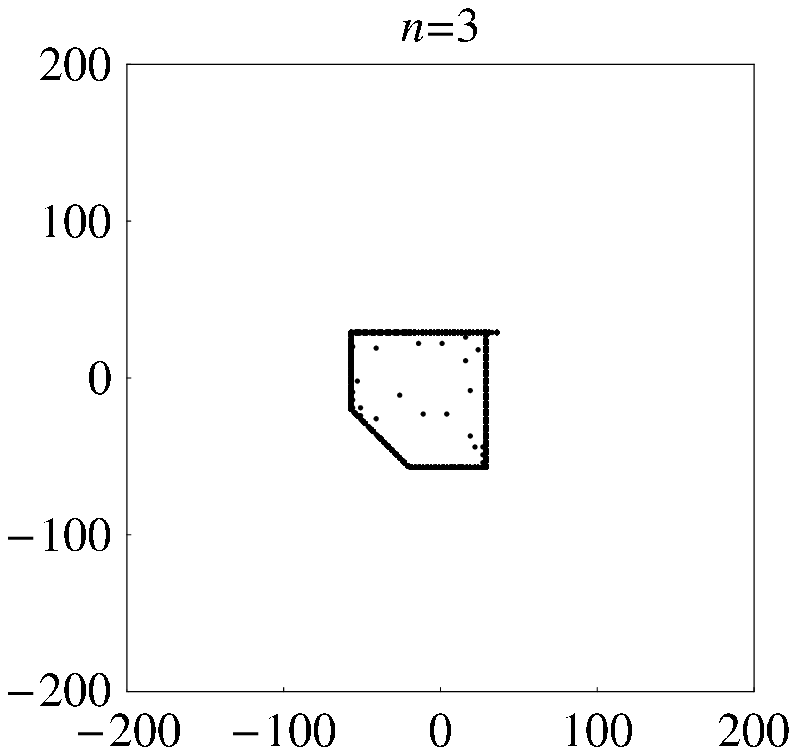} &
  \includegraphics[scale=0.5]{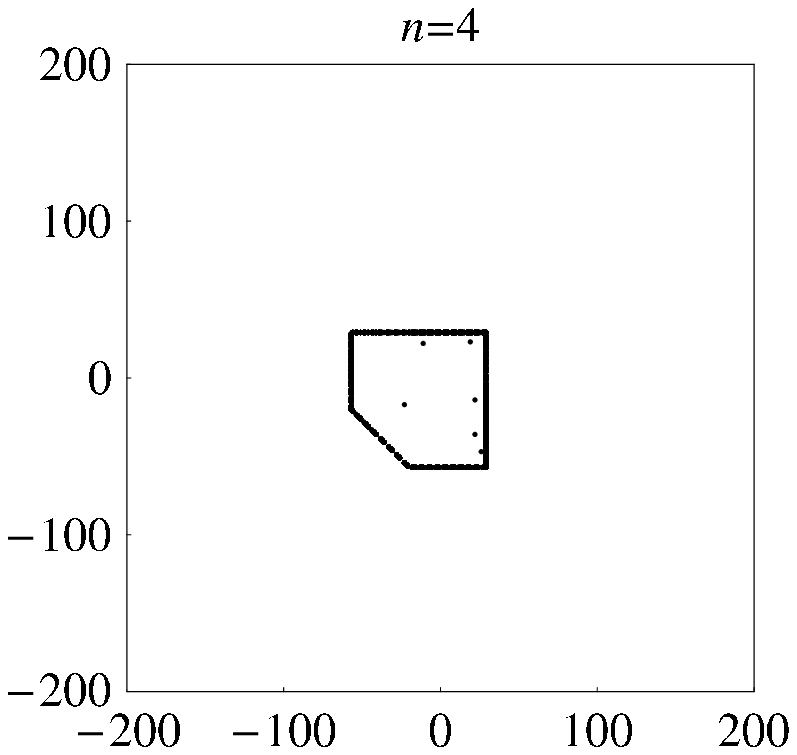} &
  \includegraphics[scale=0.5]{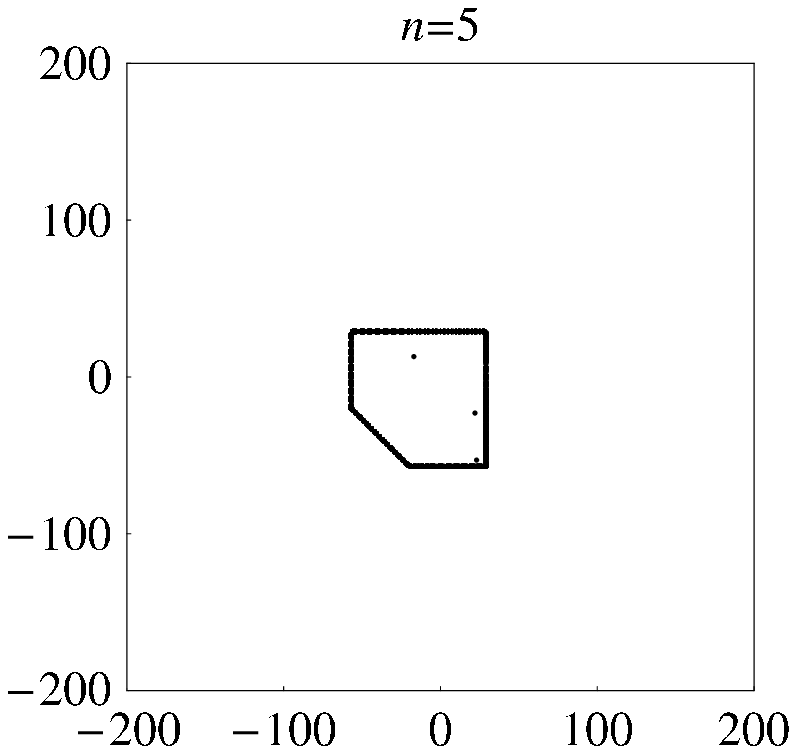} \\
  \includegraphics[scale=0.5]{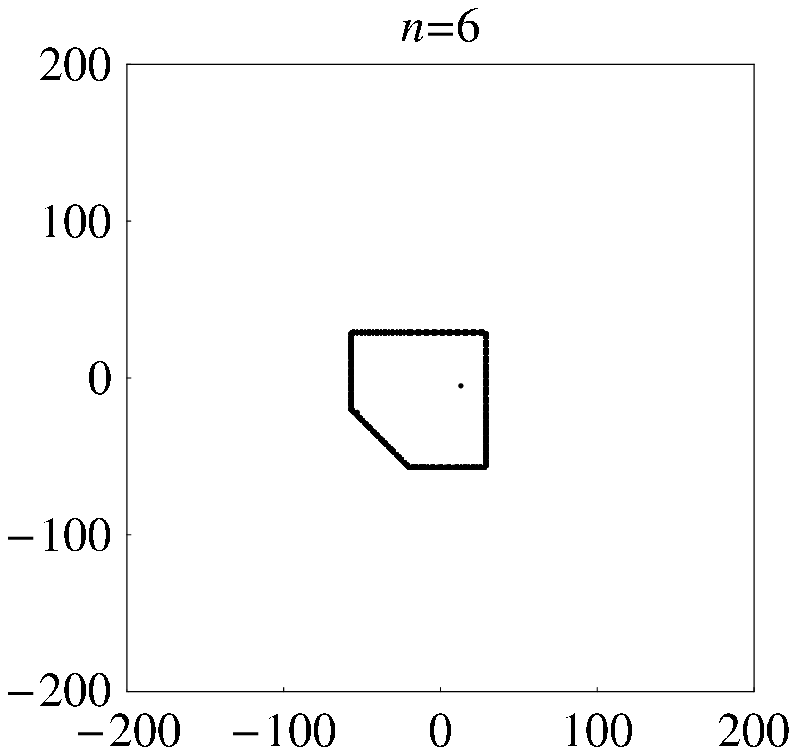} &
  \includegraphics[scale=0.5]{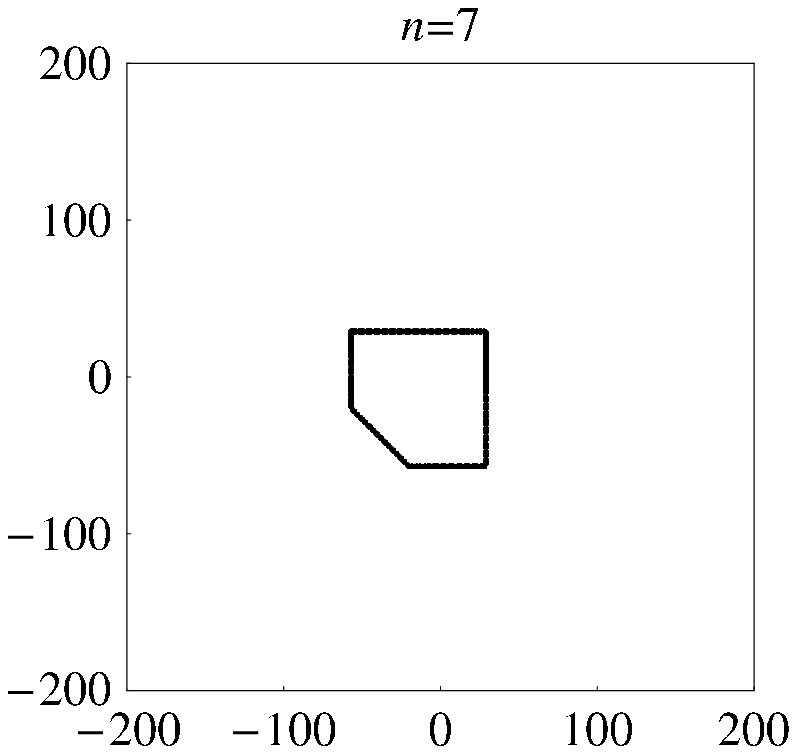} &
  \includegraphics[scale=0.5]{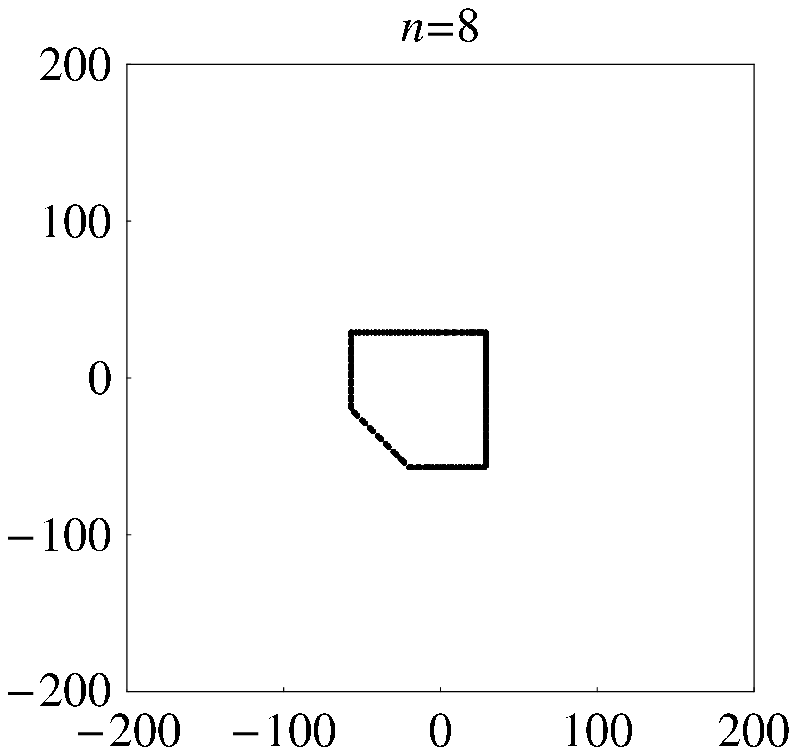}
\end{tabular}
\end{center}
\caption{Evolution of points $(x,y)=(5i,5j)$ ($-40\le i, j\le 40$) by the mapping (\ref{ultra}).  All parameters are the same as in Fig.~\ref{fig:class2 ultra1}.} \label{fig:class2 ultra2}
\end{figure}

  We use the following transformation including a new parameter $\varepsilon$ to ultradiscretize (\ref{class2}),
\begin{equation}
  a_{ij}=e^{A_{ij}/\varepsilon},\quad x_n = e^{X_n/\varepsilon},\quad h_\infty = e^{H_\infty/\varepsilon},\quad \beta_n=e^{B_n/\varepsilon}.
\end{equation}
Substituting these relations into (\ref{class2}) and taking a limit $\varepsilon\to+0$, we obtain an ultradiscrete equation,
\begin{equation}  \label{ultra}
\begin{aligned}
  X_{n+1} &= G_1(X_n)-X_{n-1}-G_3(X_n)\\
          & \qquad +\max(H_\infty+G_1(X_n),\,H_n+2X_{n-1}+G_3(X_n),\,B_n) \\
          & \qquad -\max(H_n+G_1(X_n),\,H_\infty+2X_{n-1}+G_3(X_n),\,B_n),
\end{aligned}
\end{equation}
where
\begin{equation}  \label{ultra funcs}
\begin{aligned}
  G_1(X) &= \max(A_{02}+2X,\,A_{12}+X,\,A_{22}), \\
  G_3(X) &= \max(A_{00}+2X,\,A_{01}+X,\,A_{02}), \\
  H_n &= H(X_{n-1},X_n) \\
  H(X,Y) &= \max(A_{00}+2X+2Y,\,A_{01}+X+Y+\max(X,Y),\\
         & \qquad A_{02}+2\max(X,Y),\,A_{11}+X+Y,\,A_{12}+\max(X,Y),\,A_{22}) \\
         & \qquad -X-Y.
\end{aligned}
\end{equation}
Note that $a_{ij}=0$ corresponds to $A_{ij}\to-\infty$ and the terms including $A_{ij}$ in (\ref{ultra funcs}) are eliminated from $\max$ functions.  Figure~\ref{fig:class2 ultra1} shows examples of solution to (\ref{ultra}).  Trajectory is exactly on the attractor $H_n=H_\infty$ after a finite number of time steps in both figures.  Figure~\ref{fig:class2 ultra2} shows an evolution of points in a region of the phase plane by the mapping with the same parameters in Fig.~\ref{fig:class2 ultra1}.  Points $(x, y)=(5i, 5j)$ ($-40\le i, j\le 40$) are taken as initial data.  All points in this region are mapped to the attractor after 7 steps.  Note that the only fixed point is $(-14,-14)$ and it is not shown in this figure.\par
  The polygon in the phase space defined by $H_n=H_\infty$ is always an attractor of (\ref{ultra}).  However, additional attractors of integrable trajectory often appear depending on the parameters of (\ref{ultra}).  To explain this phenomenon, let us consider a relation $a=1+b$.  This relation gives an inequality $a>b$.  If we use a transformation $a=e^{A/\varepsilon}$, $b=e^{B/\varepsilon}$, a relation between $A$ and $B$ is $A=\varepsilon\log(1+e^{B/\varepsilon})$ and $A>B$ also holds.  However, $A=\max(0,B)$ is obtained after the limit $\varepsilon\to+0$ and $A\ge B$ holds, especially $A=B$ if $B\ge0$.  This type of change of inequality often occurs in the ultradiscretization and also for the convergence condition (\ref{condition}).  It is a reason why additional attractors appear for (\ref{ultra}).
\section{Some applications}  \label{sec:application}
  We give some applications of mappings we proposed in the previous sections.  Mappings in the applications also have an explicit discrete Lyapunov function.
\subsection{Parameter adjustment}
  If a solution satisfies (\ref{condition}), $h_n$ converges to $h_\infty$ and the convergence speed depends on $\alpha_n$.  Therefore, we can easily control the local speed by adjusting $\alpha_n$.  Consider the case that $p$ in (\ref{class1 alpha}) depends on $n$.  Even in such a case, if $p$ satisfies $0<p\le1$, $h_n$ converges to $h_\infty$ monotonically.  For example, let us assume two constants $c_0$ ($0<c_0\le1$), $c_1$ and define $p$ by
\begin{equation}
  p = c_0 + (1-c_0)\sin^2(c_1(h_n-h_\infty)).
\end{equation}
\begin{figure}
\begin{center}
\begin{tabular}{cc}
  \includegraphics[scale=0.65]{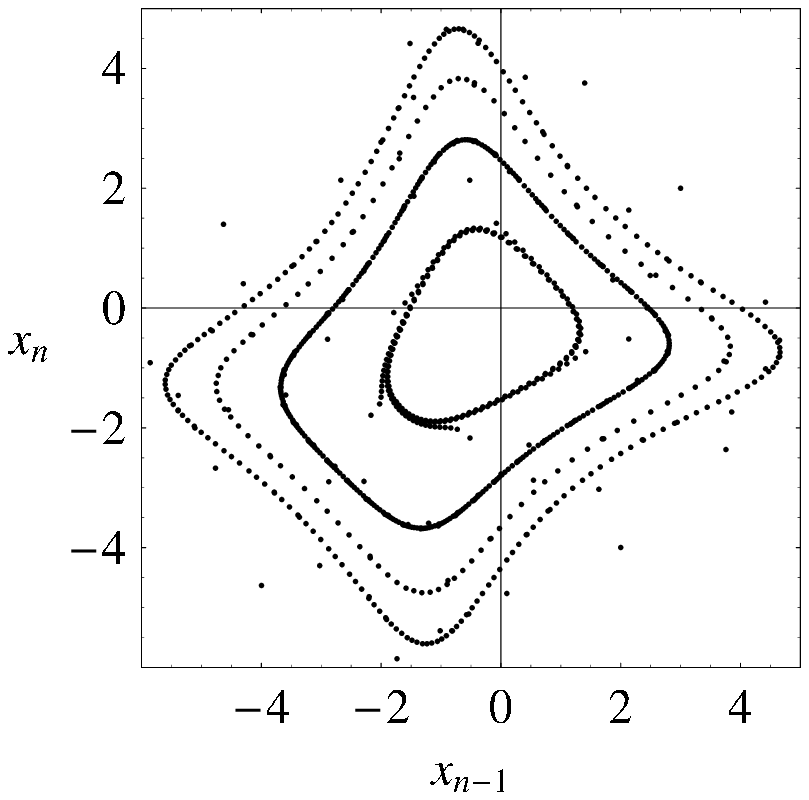} &
  \includegraphics[scale=0.8]{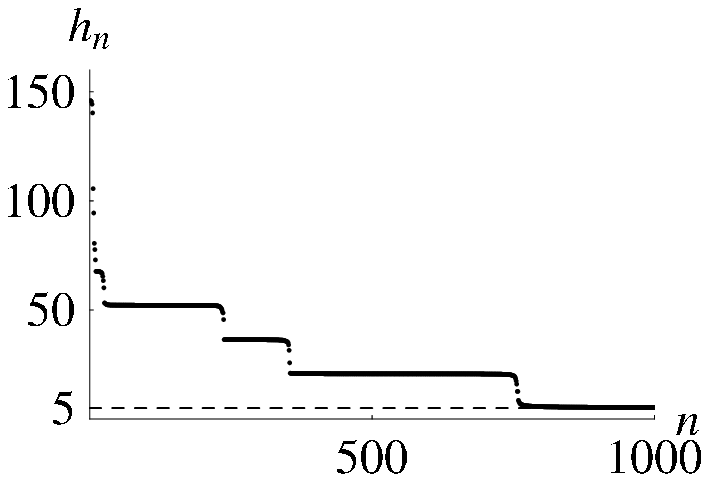} \\
  (a) & (b)
\end{tabular}
\end{center}
\caption{Mapping (\ref{class1}) with $\alpha_n$ of (\ref{class1 alpha}) adjusted by $p=c_0 + (1-c_0)\sin^2(c_1(h_n-h_\infty))$.  Parameters and initial data are the same as in Fig.~\ref{fig:class1 p=1} (a), $c_0=0.00004$ and $c_1=0.2$. (a) Solution in a phase plane, (b) Evolution of $h_n$.}
\label{fig:adjust}
\end{figure}
Figure~\ref{fig:adjust} (a) shows an example of solution in a phase plane.  The parameter $p$ depends on $h_n-h_\infty$ and $p$ is approximately equal to $c_0$ near the region where $\sin(c_1(h_n-h_\infty))\sim0$.  Therefore, if we use a small $c_0$, the convergence speed becomes slow around the region and a trajectory of solution there becomes similar to an integrable one.  Figure~\ref{fig:adjust} (b) shows an evolution of $h_n$ of the solution in Fig.~\ref{fig:adjust} (a).  It changes stepwise and converges to $h_\infty$ finally.
\subsection{Probabilistic mapping}
\begin{figure}
\begin{center}
\begin{tabular}{cc}
  \includegraphics[scale=0.65]{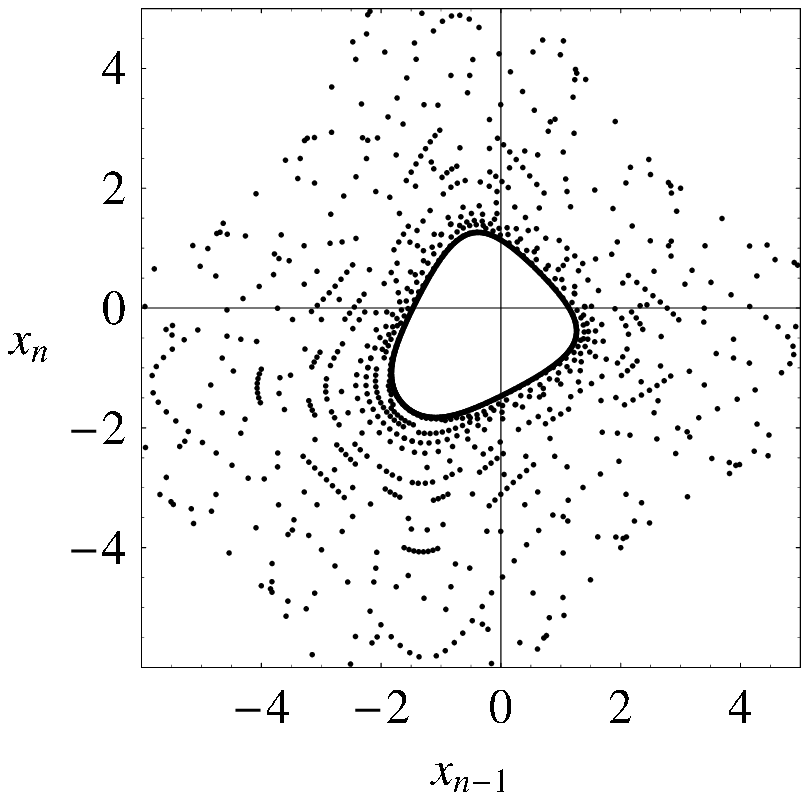} &
  \includegraphics[scale=0.8]{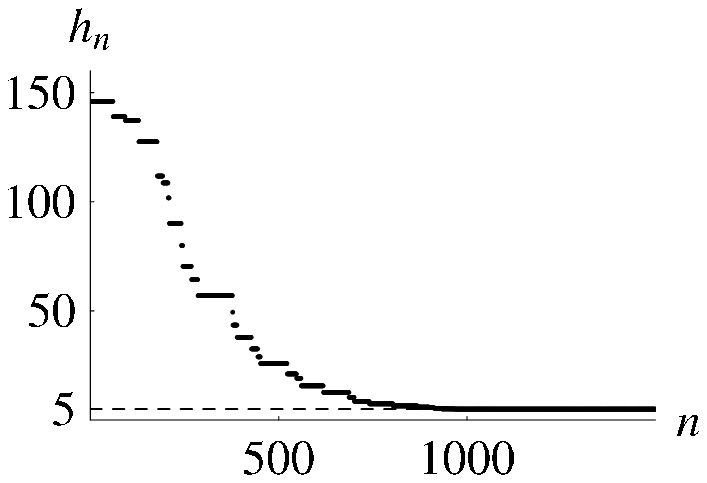} \\
  (a) & (b)
\end{tabular}
\end{center}
\caption{Probabilistic mapping (\ref{class1}) with $\alpha_n$ of (\ref{class1 alpha}) using probabilistic $p$ defined by (\ref{probability}) with $c=0.5$ and $M=30$.  Other parameters are the same as in Fig.~\ref{fig:class1 p=1} (a). (a) Solution in a phase plane, (b) Evolution of $h_n$.}
\label{fig:probability}
\end{figure}

  We can easily introduce a probabilistic parameter to $\alpha_n$ preserving the convergence.  This application is based on a similar idea to the parameter adjustment.  Consider the case that $p$ in (\ref{class1 alpha}) is a probabilistic parameter.  Even in such a case, if $p$ satisfies $0<p\le1$, $h_n$ converges to $h_\infty$ monotonically.  As an extreme example, let us assume that a value of $p$ is 0 or positive $c$ with a probability defined by
\begin{equation}  \label{probability}
  P(p=0)=1-\frac{1}{M}, \qquad P(p=c)=\frac{1}{M},
\end{equation}
with a positive integer $M$.  The mapping (\ref{class1}) becomes integrable at time steps of $p=0$ ($\alpha_n=0$) and $h_n$ becomes constant for those steps.  However, $h_n$ converges to $h_\infty$ as $n\to\infty$ since the ratio of time steps of $p=c$ is nonzero finite.  Figure~\ref{fig:probability} (a) shows a solution and Fig.~\ref{fig:probability} (b) shows an evolution of $h_n$.
\subsection{Coupled mappings}
  The solution $x_n$ to (\ref{class2}) satisfying (\ref{class2:condition}) is always positive when positive initial values are used.  Moreover, the only condition required for $\beta_n$ in (\ref{class2:condition}) is $\beta_n\ge0$.  Therefore we can easily couple some of mappings (\ref{class2}).  The general form of a system of $x_n^{(1)}$, $x_n^{(2)}$, $\cdots$, $x_{n}^{(M)}$ is defined by
\begin{equation}
\begin{aligned}
  & x_{n+1}^{(j)} = \frac{g_1^{(j)}(x_n^{(j)})}{x_{n-1}^{(j)}g_3^{(j)}(x_n^{(j)})}\\
\times& \frac{h_\infty^{(j)} g_1^{(j)}(x_n^{(j)})+h_n^{(j)}(x_{n-1}^{(j)})^2g_3^{(j)}(x_n^{(j)})+\beta^{(j)}(n,x_{n-1}^{(1)},\cdots,x_{n-1}^{(M)},x_n^{(1)},\cdots,x_n^{(M)})}{h_n^{(j)}g_1^{(j)}(x_n^{(j)})+h_\infty^{(j)} (x_{n-1}^{(j)})^2g_3^{(j)}(x_n^{(j)})+\beta^{(j)}(n,x_{n-1}^{(1)},\cdots,x_{n-1}^{(M)},x_n^{(1)},\cdots,x_n^{(M)})},
\end{aligned}
\end{equation}
where
\begin{equation}
  h_n^{(j)} = h(x_{n-1}^{(j)}, x_n^{(j)}).
\end{equation}
Note that $g_1^{(j)}$, $g_3^{(j)}$, $h_\infty^{(j)}$ and $\beta^{(j)}$ can be defined independently for every $j$ and $\beta^{(j)}$ is an arbitrary positive function.\par
\begin{figure}
\begin{center}
\begin{tabular}{cc}
  \includegraphics[scale=0.75]{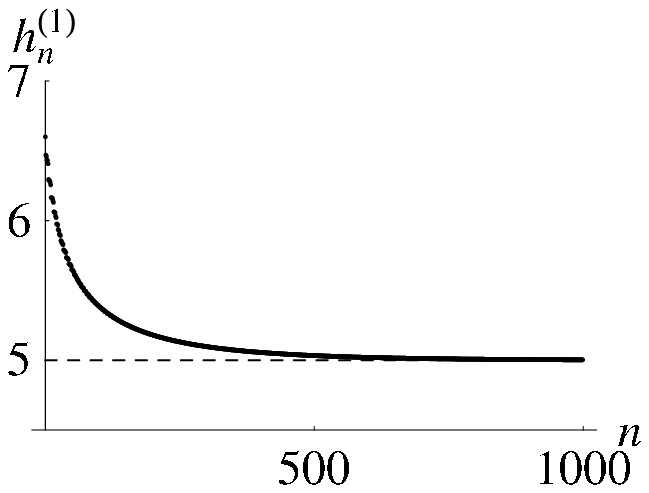} &
  \includegraphics[scale=0.75]{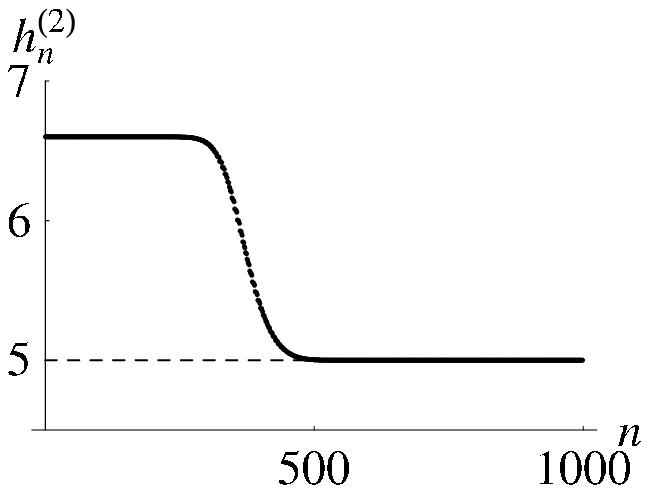} \\
  (a) & (b)
\end{tabular}
\end{center}
\caption{Evolution of the discrete Lyapunov function (a) $h_n^{(1)}$ and (b) $h_n^{(2)}$.  Common parameters to define $g_1^{(j)}$, $g_3^{(j)}$, $h_n^{(j)}$ are $(a_{00},a_{01}, a_{02}, a_{11}, a_{12}, a_{22}) = (1, 0, 0, 0, 1, 2)$, $h_\infty^{(1)}=h_\infty^{(2)}=5$, $\beta^{(1)}=1000$, $\beta^{(2)}=\exp(100(h_n^{(1)} - h_\infty^{(1)}))$, $x_0^{(1)}=x_0^{(2)}=5$ and $x_1^{(1)}=x_1^{(2)}=1$.}
\label{fig:couple}
\end{figure}
\begin{figure}
\begin{center}
\begin{tabular}{cc}
  \includegraphics[scale=0.7]{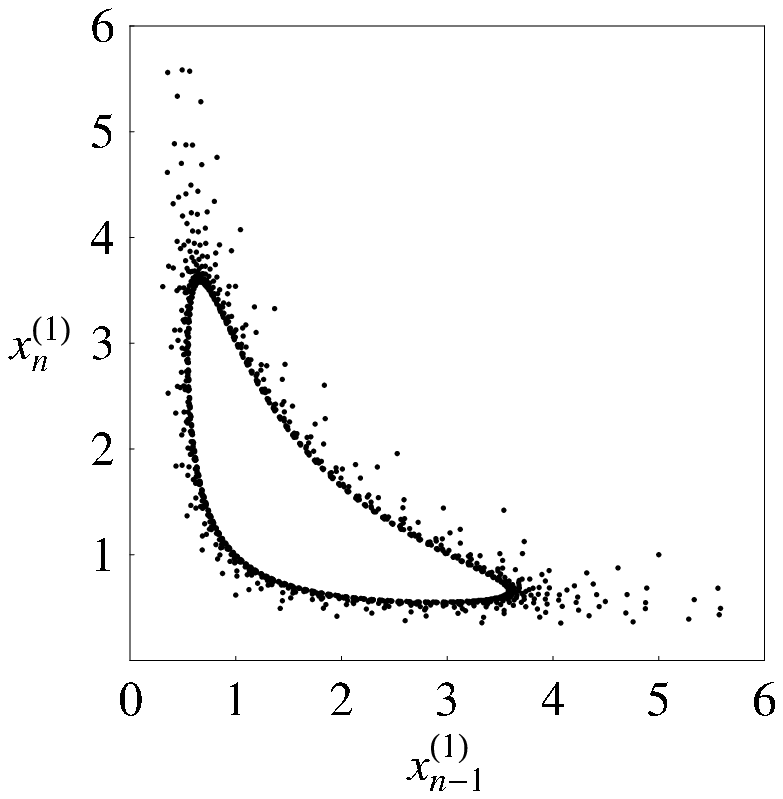} &
  \includegraphics[scale=0.7]{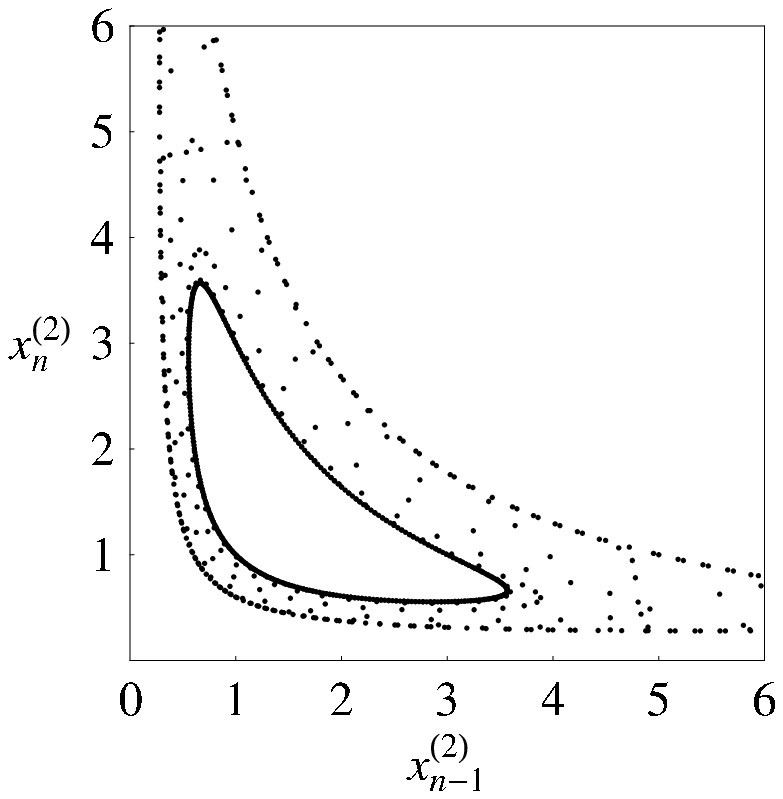} \\
  (a) & (b)
\end{tabular}
\end{center}
\caption{Solutions in a phase plane for (a) $x_n^{(1)}$ and (b) $x_n^{(2)}$.  Mappings and initial data are the same as in Fig.~\ref{fig:couple}.}
\label{fig:couple2}
\end{figure}

  Figure~\ref{fig:couple} shows a simple case, coupled two mappings ($M=2$) where
\begin{equation}
\begin{aligned}
  &g_1^{(1)}(x)=g_1^{(2)}(x),\qquad g_3^{(1)}(x)=g_3^{(2)}(x),\\
  &h^{(1)}(x,y)=h^{(2)}(x,y),\qquad h_\infty^{(1)}=h_\infty^{(2)}, \\
  &\beta^{(1)}=1000,\qquad \beta^{(2)}=\exp(100(h_n^{(1)} - h_\infty^{(1)})).
\end{aligned}
\end{equation}
In this case, $x_n^{(1)}$ is not affected by $x_n^{(2)}$, but $x_n^{(2)}$ is by $x_n^{(1)}$ through $h_n^{(1)}$.  If initial data of $x_n^{(1)}$ satisfy $h_0^{(1)} \gg h_\infty^{(1)}$, then $\beta^{(2)}\gg0$ and the mapping for $x_n^{(2)}$ is similar to an integrable one initially.  Therefore the convergence speed of $h_n^{(2)}$ to $h_\infty^{(2)}$ is slow initially.  However, $x_n^{(1)}$ satisfies $h_n^{(1)}\sim h_\infty^{(1)}$ for $n\gg0$, then $\beta^{(2)}$ becomes small and the convergence speed of $h_n^{(2)}$ becomes larger.  Figure~\ref{fig:couple} (a) and (b) show an evolution of $h_n^{(1)}$ and of $h_n^{(2)}$ respectively.  Figure~\ref{fig:couple2} (a) and (b) show solutions $x_n^{(1)}$ and $x_n^{(2)}$ in a phase plane respectively.
\section{Summary and concluding remarks}  \label{sec:remarks}
  We propose second-order discrete mappings as an extension of the QRT mappings.  They have an explicit discrete Lyapunov function and the asymptotic behavior of a solution is clearly understood.  The function is the same as the conserved quantity of the corresponding QRT mapping.  This fact suggests that we can extend an integrable system to nonintegrable one together with its characteristic structure.\par
  Moreover, a differential and an ultradiscrete equations corresponding to the mapping can be derived by taking a limit of a parameter.  Both equations have also an explicit Lyapunov function.  Thus we give a set of discrete, differential and ultradiscrete equations of which behavior of solutions obey the same structure.  There are many examples of this kind of set among integrable equations.  It suggests that there may be some mathematical structure in the equations proposed.  Note that although it is not so difficult to construct a differential equation with an explicit Lyapunov function, it is difficult to discretize or ultradiscretize the equation together with its Lyapunov function.\par
  We show some applications of the discrete mappings.  We give a mapping with an adjusted parameter, a probabilistic mapping and coupled mappings preserving the discrete Lyapunov function.  As shown in the previous section, the existence of the function makes it possible to construct applicable systems of which behavior is easy to control.\par
  Lastly we list below some future problems.
\begin{itemize}
\item
  It is clear that the mapping proposed is nonintegrable since they are not reversible.  However, more detailed study may reveal a relation between the mapping and its integrable correspondence.
\item
  There may be another characteristic quantity of the mappings.  Utilizing the quantity together with the Lyapunov function, we may be able to give more information about solutions.
\item
  Though we concentrate on modifying the QRT mapping, a similar procedure may be applied to other integrable mappings.  Generalizing the method used is an interesting problem.
\item
  We propose the applications of mappings only as demonstrative examples.  We should find an application as a model of more realistic phenomenon.
\end{itemize}
\ack
  The authors express our sincere thanks to Prof. Ryogo Hirota for his advices on the derivation of differential equations and to Prof. Yutaka Ishii for his significant comments on dynamical systems.

\end{document}